\documentclass[aps,pra,twocolumn]{revtex4-1}

\usepackage{bm,bbm}
\usepackage{amsmath,amsthm,amssymb,amsbsy}
\usepackage{graphicx,color}
\usepackage{amsfonts}
\usepackage{amssymb}
\usepackage{times,txfonts}
\usepackage{hyperref}

\newcommand{\ket}[1]{|#1\rangle}
\newcommand{\bra}[1]{\langle#1|}

\begin{document}

\title{Observation of Time-Invariant Coherence in a Nuclear Magnetic Resonance Quantum Simulator}

\author
{Isabela~A.~Silva,$^{1,2}$ Alexandre~M.~Souza,$^{3}$ Thomas~R.~Bromley,$^{2}$
Marco~Cianciaruso,$^{2,4}$
Raimund~Marx,$^5$
Roberto~S.~Sarthour,$^{3}$ Ivan~S.~Oliveira,$^{3}$
Rosario~Lo~Franco,$^{1,2,6,7}$
Steffen~J.~Glaser,$^5$
Eduardo~R.~deAzevedo,$^{1}$ Diogo O.~Soares-Pinto,$^{1}$
and Gerardo Adesso$^{2,\ast}$  \vspace*{.2cm}}
\affiliation{{$^{1}$Instituto de F\'isica de S\~{a}o Carlos, Universidade de S\~{a}o Paulo,}
{CP 369, 13560-970, S\~{a}o Carlos, SP, Brazil}\\
{$^{2}\mbox{School of Mathematical Sciences, The University of Nottingham,
University Park, Nottingham NG7 2RD, United Kingdom}$}\\
{$^{3}$Centro Brasileiro de Pesquisas F\'isicas, Rua Dr. Xavier Sigaud 150,}
{22290-180 Rio de Janeiro,  RJ, Brazil}\\
{$^{4}$Dipartimento di Fisica ``E. R. Caianiello'', Universit\`a degli Studi di Salerno,}
{Via Giovanni Paolo II, I-84084 Fisciano (SA), Italy}\\
{$^{5}$Department of Chemistry, Technische Universit\"at M\"unchen, Lichtenbergstr.~4, 85747, Garching, Germany} \\
{$^6$Dipartimento di Energia, Ingegneria dell'Informazione e Modelli Matematici, Universit\`a di Palermo, Viale delle Scienze, Edificio 9, 90128 Palermo, Italy}  \\ {$^7$Dipartimento di Fisica e Chimica, Universit\`a di Palermo, Via Archirafi 36, 90123 Palermo, Italy} \\
{$^\ast$To whom correspondence should be addressed; E-mail: gerardo.adesso@nottingham.ac.uk}
}

\date{September 19, 2016}

\begin{abstract}
The ability to live in coherent superpositions is a signature trait of quantum systems and constitutes an irreplaceable resource for quantum-enhanced technologies. However, decoherence effects usually destroy quantum superpositions. It has been recently predicted that, in a composite quantum system exposed to dephasing noise, quantum coherence in a transversal reference basis can stay protected for indefinite time. This can occur for a class of quantum states independently of the measure used to quantify coherence, and requires no control on the system during the dynamics. Here, such an invariant coherence phenomenon is observed experimentally in two different setups based on nuclear magnetic resonance at room temperature, realising an effective quantum simulator of two- and four-qubit spin systems. Our study further reveals a novel interplay between coherence and various forms of correlations, and highlights the natural resilience of quantum effects in complex systems.
\end{abstract}

\maketitle

Successfully harnessing genuine nonclassical effects is predicted to herald a new wave of technological devices with a disruptive potential to supersede their conventional counterparts \cite{Dowling2003}. This prediction is now coming of age, and an international race is on to translate the power of quantum technologies into commercial applications to networked communication, computing, imaging, sensing and simulation \cite{Money}. Quantum {\em coherence} \cite{AlexRMP}, which incarnates the wavelike nature of matter and the essence of quantum parallelism \cite{Leggett1980}, is the primary  ingredient enabling a supraclassical performance in a wide range of such applications. Its key role in quantum algorithms, optics, metrology, condensed matter physics,  and nanoscale thermodynamics is actively investigated and widely recognised \cite{Glauber1963,Braunstein1994,Giovannetti2004,Fahmy2008,Lostaglio2015,Hillery2015,Ma2015,Matera2015,Napoli2016}.  Furthermore, coherent quantum effects have been observed in large molecules \cite{Arndt} and are advocated to play a functional role in even larger biological complexes \cite{Lloyd2011, Engel2007, Panitchayangkoon2010,   Chin2013}. However, coherence is an intrinsically fragile property which typically vanishes at macroscopic scales of space, time, and temperature: the disappearance of coherence, i.e.~{\it decoherence} \cite{Zurek2003}, in quantum systems exposed to environmental noise is one of the major hindrances still threatening the scalability of most quantum machines. Numerous efforts have been thus invested in recent years into devising feasible control schemes to preserve coherence in open quantum systems \cite{Roadmap}, with notable examples including dynamical decoupling \cite{DynDec,DynDecAMS}, quantum feedback control \cite{FeedControl} and error correcting codes \cite{Nielsen2010}.

\begin{figure*}[t!]
\centering
\vspace*{-.5cm}
\begin{minipage}[b]{.655\textwidth}
\vspace*{-1cm}
\includegraphics[width=\textwidth]{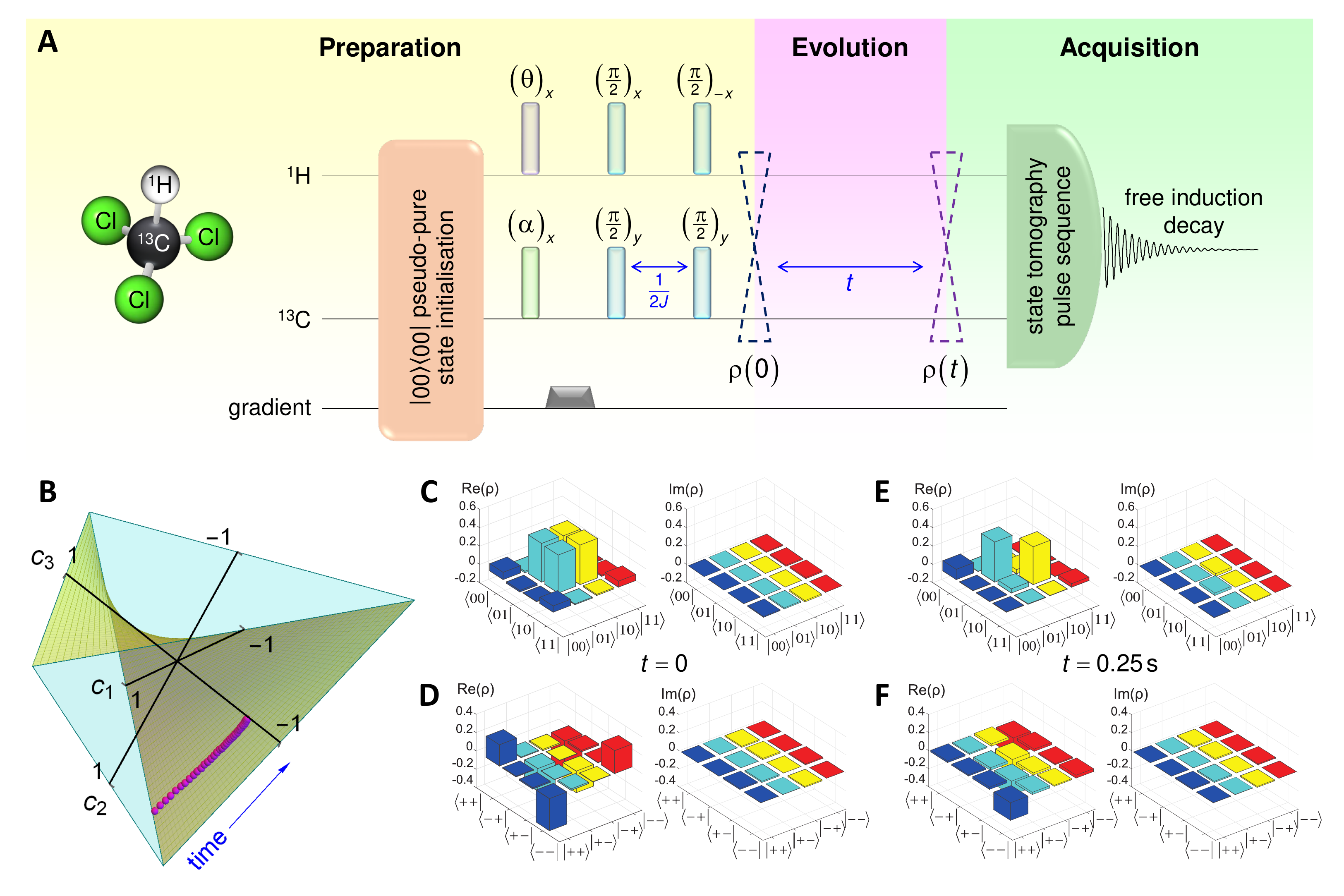} \\[-0.1cm]
\includegraphics[width=.98\textwidth]{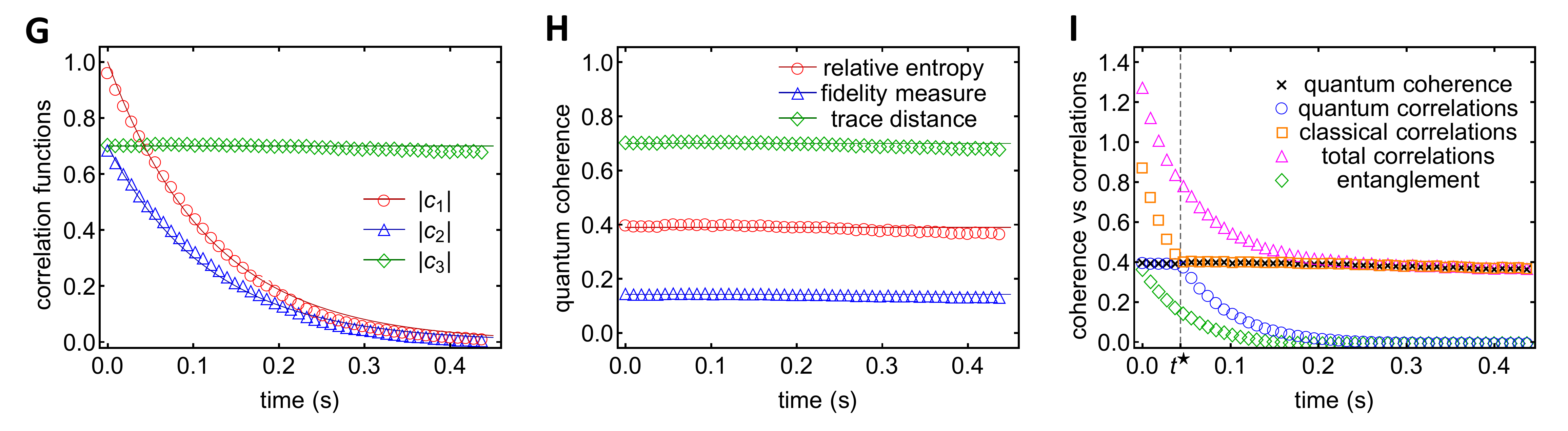}
\end{minipage}
\begin{minipage}[b]{.34\textwidth}
\caption{\footnotesize {\sf \bfseries A}: Pulse sequence to prepare two-qubit BD states encoded in the  $^1$H and $^{13}$C nuclear spins of Chloroform; the rf pulse $(\theta)_\mu$ realises a qubit rotation by $\theta$ about the spin-$\mu$ axis, $J$ is the scalar spin-spin coupling, and time flows from left to right. {\sf \bfseries B}: Dynamics of the experimental states $\rho(t)$ (magenta points) in the space of the spin-spin correlation triple $c_j = \langle \sigma_j \otimes \sigma_j\rangle$, $j=1,2,3$; all BD states (\ref{eq:BDstate}) fill the light blue tetrahedron, while the subclass of states spanning the inscribed green surface are predicted to have time-invariant coherence in the plus/minus basis according to any measure of Eq.~(\ref{CD}) \cite{Frozen}. {\sf \bfseries C}--{\sf \bfseries F}: full tomographies of the experimental states as prepared at time $t=0$ ({\sf \bfseries C}, {\sf \bfseries D}) and after $t = 0.25$ s of free evolution   ({\sf \bfseries E}, {\sf \bfseries F}), recorded in the computational basis (top row) and in the plus/minus basis (bottom row).
{\sf \bfseries G}: Evolution of (absolute values of) the correlation functions $|c_j|$ in the experimental states (points), along with  theoretical predictions (solid lines) based on phase damping noise with our measured relaxation times. {\sf \bfseries H}: Experimental observation of time-invariant  coherence (in the plus/minus basis), measured by relative entropy (red circles) \cite{Baumgratz2014}, fidelity-based measure (blue triangles) \cite{Streltsov2015}, and normalised trace distance (green diamonds) \cite{Frozen}, equal to  $l_1$ norm \cite{Baumgratz2014} in BD states. The slight negative slope is due to the subdominant effect of amplitude damping.
{\sf \bfseries I}: Experimental dynamics of coherence and all forms of correlations \cite{Modi2012} measured via  relative entropies (theoretical curves omitted for graphical clarity).
In panels {\sf \bfseries G}--{\sf \bfseries I}, experimental errors due to small pulse imperfections (0.3\% per pulse) result in error bars within the size of the data points.}
\label{figplan}
\end{minipage}
\end{figure*}


In this Letter we demonstrate a fundamentally different mechanism. We observe experimentally that quantum coherence in a composite system, whose subsystems are all affected by decoherence, can remain {\it de facto} invariant for arbitrarily long time without any external control. This phenomenon was recently predicted to occur  for a particular family of initial states of  quantum systems of any (however large) even number of qubits \cite{Frozen}, and is here demonstrated in a room temperature liquid-state nuclear magnetic resonance (NMR) quantum simulator \cite{Knill98,Marx2000,Sharf2000,marx,livro_ivan} with two different molecules, encompassing two-qubit and four-qubit spin ensembles. After initialisation into a so-called generalised Bell diagonal state \cite{Frozen}, the multiqubit ensemble is left to evolve under naturally occurring phase damping noise. Constant coherence in a reference basis (transversal to the noise direction) is then observed within the experimentally considered timescales up to the order of a second. Coherence is measured according to a variety of recently proposed quantifiers \cite{Baumgratz2014}, and its permanence is verified to be measure-independent. We also reveal how coherence  captures quantitatively a dynamical interplay between classical and general quantum correlations \cite{Mazzola2010}, while any entanglement may rapidly disappear \cite{Yu2009}. For more general initial states, we prove theoretically that coherence can decay yet remains above a guaranteed threshold at any time, and we observe this experimentally in the two-qubit instance.
The present study advances our physical understanding of the resilience of quantum effects against decoherence.

Quantum coherence manifests when a quantum system is in a superposition of multiple states taken from a reference basis. The reference basis can be indicated by the physics of the problem under investigation (e.g.~one may focus on the energy eigenbasis when addressing coherence in transport phenomena and thermodynamics) or by a task for which coherence is required (e.g.~the estimation of a magnetic field in a certain direction). Here, for an $N$-qubit system, having in mind a magnetometry setting \cite{KoloPRX}, we can fix the reference basis to be the `plus/minus' basis $\{\ket{\pm}^{\otimes N}\}$, where $\{\ket{\pm}\}$ are the eigenstates of the $\sigma_1$ Pauli operator, which describes the $x$ component of the spin on each qubit \cite{Nielsen2010}. Any state with density matrix $\delta$ diagonal in the plus/minus basis will be referred to as incoherent. According to a recently formulated resource theory \cite{Baumgratz2014,Streltsov2015,Winter2016,AlexRMP}, the degree of quantum coherence in the state $\rho$ of a quantum system can be quantified in terms of how distinguishable  $\rho$ is from the set of incoherent states,
\begin{equation}\label{CD}
C_D(\rho)=\inf_{\mbox{$\delta$ incoherent}} D(\rho, \delta)\,,
\end{equation}
where the distance $D$ is assumed jointly convex and contractive under quantum channels, as detailed in the Supplemental Material \cite{EPAP}. \nocite{Brandao2015,Vedral1997,Modi2010,Bromley2014,Paula2014,Aaronson2013,Cornelio2012,Cianciaruso2015Entanglement} In general, different measures of coherence induce different orderings on the space of quantum states, as it happens e.g.~for entanglement or other resources. A consequence of this is that, for states of a single qubit, it is impossible to find a nontrivial noisy dynamics under which coherence is naturally preserved when measured with respect to all possible choices of $D$ in Eq.~(\ref{CD}). As predicted in \cite{Frozen}, such a counterintuitive situation can occur instead for larger composite systems.  Here we observe this phenomenon experimentally.

Our NMR setup realises an effective quantum simulator, in which $N$-qubit states  $\rho$ can be prepared by manipulating the deviation matrix from the thermal equilibrium density operator of a spin ensemble \cite{Knill98,Sharf2000}, via application of radiofrequency (rf) pulses and evolution under spin interactions \cite{Nielsen2010,livro_ivan}.  The scalability of the setup relies on availability of suitably large controllable molecules in liquid-state solutions.

We first encoded a two-qubit system in a Chloroform (CHCl$_3$) sample enriched with $^{13}$C, where the $^1$H and $^{13}$C spin-$\frac12$ nuclei are associated to the first and second qubit, respectively. This experiment was performed in a Varian 500 MHz liquid-NMR spectrometer at room temperature, according to the plan illustrated in Fig.~\ref{figplan}{\sf  A}.
The state preparation stage allowed us to initialise the system in any state obtained as a mixture of maximally entangled Bell states, that is, any Bell diagonal (BD) state \cite{Paula2013}. These states take the form
 \begin{equation}\label{eq:BDstate}
\rho=\mbox{$\frac{1}{4}\left(\mathbb{I}  \otimes \mathbb{I} + \sum_{j=1}^3 c_j\ \sigma_j \otimes\sigma_j\right)$}\,,
\end{equation}
where $\{\sigma_j\}$ are the Pauli matrices and  $\mathbb{I}$ is the identity operator on each qubit; they are completely specified by the spin-spin correlation functions $c_j = \langle \sigma_j \otimes \sigma_j \rangle$ for $j=1,2,3$, and can be conveniently represented in the space spanned by these three parameters as depicted in Fig.~\ref{figplan}{\sf  B}. We aimed to prepare specifically a BD state with initial correlation functions $c_1(0) = 1$, $c_2(0) = 0.7$ and $c_3(0) = -0.7$, by first initialising the system in the pseudo-pure state $\ket{00}\bra{00}$ as described in Refs.~\cite{Nielsen2010, Knill98}, and then implementing the sequence of rf pulses shown in Fig.~\ref{figplan}{\sf  A} with $\theta = \pi$ and $\alpha = \arccos(-0.7) \approx 134^{\circ}$.

After state preparation, the system was allowed to evolve freely during a period of time $t$, with $t$ increased for each trial in increments of $2/J$ from $0$ to $0.5$ s  (where $J \approx 215$ Hz is the scalar spin-spin coupling constant \cite{EPAP}), in order to obtain the complete dynamics. In the employed setup, the two main sources of decoherence can be modelled as Markovian phase damping and generalised amplitude damping channels acting on each qubit, with characteristic relaxation times $T_2$ and $T_1$, respectively \cite{EPAP}. For our system, the relaxation times were measured as $T_1^H = 7.53$ s, $T_2^H = 0.14$ s, $T_1^C = 12.46$ s, $T_2^C = 0.90$ s which implies that ${T}_1^{H,C} \gg {T}_2^{H,C}$. Therefore, considering also the time domain of the experiment, only the phase damping noise can be seen to have a dominant effect. 

The final stage consisted of performing full quantum state tomography for each interval of time $t$, following the procedure detailed in \cite{EPAP,LongQST}. Instances of the reconstructed experimental states at $t=0$ and $t=0.25$ s are presented in Fig.~\ref{figplan}{\sf  C}--{\sf  F}. The fidelity of the initial state with the ideal target was measured at $99.1\%$, testifying the high degree of accuracy of our preparation stage. We verified that the evolved state remained of the BD form (\ref{eq:BDstate}) during the whole dynamics with fidelities above $98.5\%$: we could then conveniently visualise the dynamics focusing on the evolution of the spin-spin correlation triple $\{c_j(t)\}$, as indicated by magenta points in Fig.~\ref{figplan}{\sf  B}. The time evolution of the triple $\{c_j(t)\}$ is detailed in Fig.~\ref{figplan}{\sf  G}.

\begin{figure}[b!]
\centering
\includegraphics[width=0.465\textwidth]{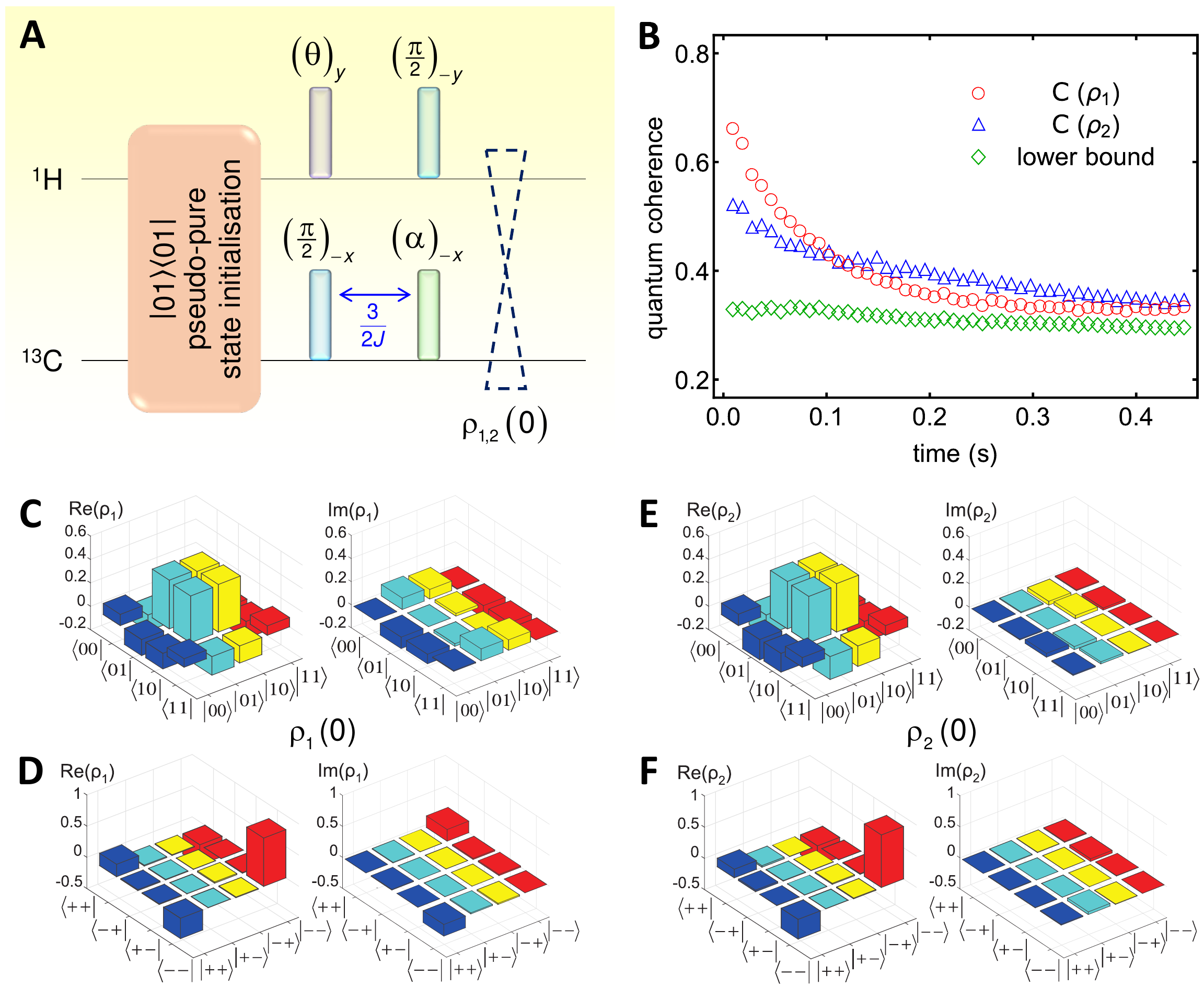}
\caption{\label{fignonbd} \footnotesize {\sf \bfseries A}: Modified preparation stage to engineer non-BD two-qubit states. We prepared two states $\rho_1$ and $\rho_2$ with purity $0.92$ and $0.93$ respectively, setting phases $\theta \approx 0.94$ rad, $\alpha = \pi/3$ for $\rho_1$, and $\theta \approx 0.78$ rad, $\alpha = \pi/2$ for  $\rho_2$. The evolution and acquisition stages were as in Fig.~\ref{figplan}{\sf A}. Full tomographies of the produced states at  $t=0$ are presented in:  {\sf \bfseries C} ($\rho_1$, computational basis), {\sf \bfseries D} ($\rho_1$, plus/minus basis), {\sf \bfseries E} ($\rho_2$, computational basis), and {\sf \bfseries F} ($\rho_2$, plus/minus basis).  {\sf \bfseries B}: Dynamics of the relative entropy of coherence in the prepared states, along with the lower bound inferred from the evolution of their spin-spin correlation functions. The experimental errors are estimated as in Fig.~\ref{figplan}.\vspace*{-.2cm}}
\end{figure}

From the acquired state tomographies during the relaxation progress, we measured the dynamics of quantum coherence in our states adopting all the known geometric coherence monotones proposed in the literature, as shown in Fig.~\ref{figplan}{\sf  H}.  All quantifiers were found simultaneously constant within the experimental confidence levels, revealing a universal resilience of quantum coherence in the dynamics under investigation.
Note that the observed time-invariant coherence is a nontrivial feature which only occurs under particular dynamical conditions. A theoretical analysis \cite{Frozen,EPAP} predicts in fact that, for all BD states evolving such that their spin-spin correlations obey the condition $c_2(t) = -c_1(t) c_3(t)$ (corresponding to the lime green surface in Fig.~\ref{figplan}{\sf  B}), any valid measure of coherence as defined in Eq.~(\ref{CD}) with respect to the plus/minus basis should remain constant at any time $t$.
As evident from the placement of the data points in Fig.~\ref{figplan}{\sf  B}, our setup realised precisely the predicted dynamical conditions for time-invariant coherence, with no further control during the relaxation. Our experiment thus demonstrated a nontrivial spontaneous occurrence of long-lived quantum coherence under Markovian dynamics.

We remark that the observed effect is distinct from the physical mechanism of long-lived singlet states also studied in NMR \cite{Long-lived}, and from an instance of decoherence-free subspace \cite{Dfsexp}. In the latter case, an open system dynamics can act effectively as a unitary evolution on a subset of quantum states, automatically preserving their entropy and other informational properties. In our case, the states are instead degraded with time, but only their coherence in the considered reference basis remains unaffected. We verified this by measuring other indicators of correlations \cite{Modi2012} in our states as a function of time. Fig.~\ref{figplan}{\sf I} shows the dynamics of entanglement, classical, quantum, and total correlations (defined in \cite{EPAP}), as well as coherence. While entanglement is found to undergo a sudden death \cite{Yu2009,Almeida2007} at $\approx 0.21$ s, a sharp transition between the decay of classical and quantum correlations is observed at the switch time $t^\star =\frac{T_2^H T_2^C}{T_2^H+T_2^C} \ln\left\vert \frac{c_1(0)}{c_3(0)}\right\vert \approx 0.043$ s. Such a puzzling feature has been reported earlier theoretically \cite{Mazzola2010,Cianciaruso2015} and experimentally \cite{Xu2010,Auccaise2011,Paula2013}, but here we reveal the prominent role played by coherence in this dynamical picture. Namely, coherence in the plus/minus basis is found equal to quantum correlations before $t^\star$ and to classical ones after $t^\star$, thus remaining constant at all times. This novel interplay between coherence and correlations, observed in our natural decohering conditions, is expected to manifest for any valid choice of geometric quantifiers used to measure the involved quantities \cite{Frozen}; for instance, in Fig.~\ref{figplan}{\sf I} we picked all measures based on relative entropy.

One might wonder how general the reported phenomena are if the initial states differ from the BD states of Eq.~(\ref{eq:BDstate}). In \cite{EPAP} we prove that, given an arbitrary state $\rho$ with spin-spin correlation functions $\{c_j\}$, its coherence with respect to any basis is always larger than the coherence of the generalised BD state defined by the same correlation functions.
This entails that, even if coherence in arbitrary states may decay under noise, it will stay above a threshold guaranteed by the coherence of corresponding BD states. To demonstrate this, we modified our preparation scheme to engineer more general two-qubit states (Fig.~\ref{fignonbd}{\sf A}). We prepared two different pseudo-pure states $\rho_1$ and $\rho_2$, both with matching initial correlation triple $c_1(0) = 0.95$, $c_2(0) = 0.62$, $c_3(0) = -0.65$,  within the experimental accuracy. We then measured their coherence  dynamics under natural evolution as before. For both of them, we observed a decay of coherence (albeit with different rates) towards a common time-invariant lower bound, which
was determined solely by the evolution of the spin-spin correlation functions, regardless of the specifics of the states (Fig.~\ref{fignonbd}{\sf B}).

\begin{figure*}[t!]
\centering
\vspace*{-.5cm}
\includegraphics[width=0.644\textwidth]{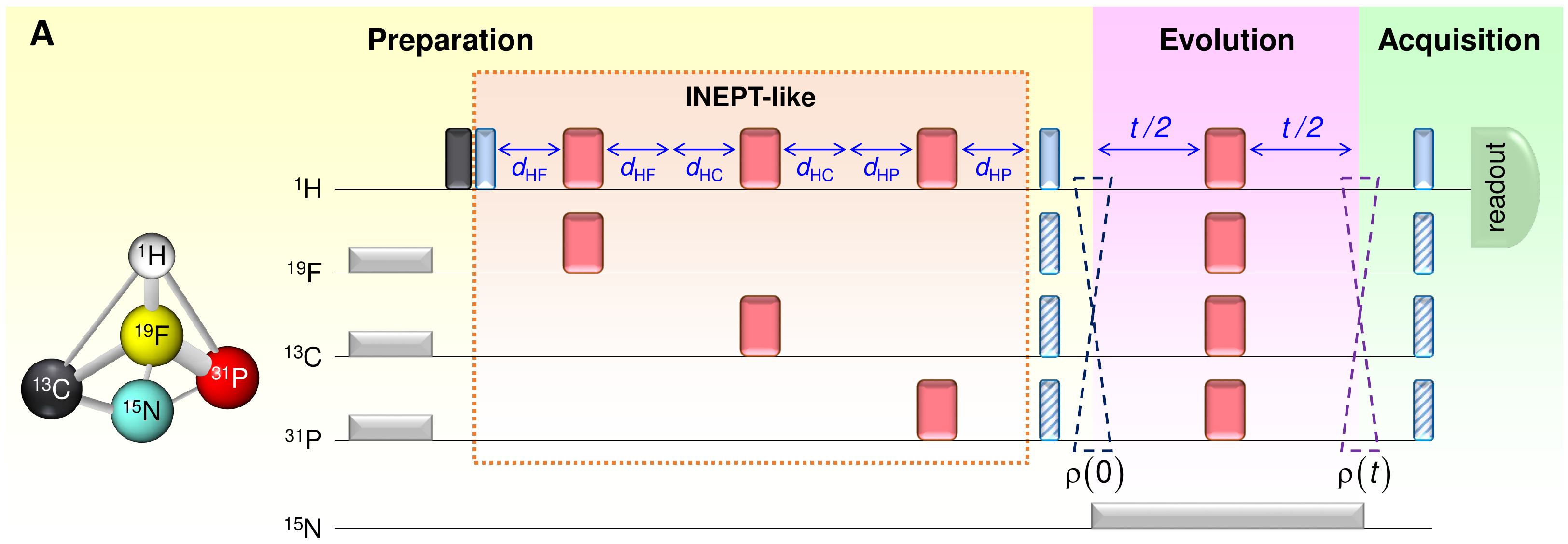} \hspace*{.1cm}
\includegraphics[width=0.164\textwidth]{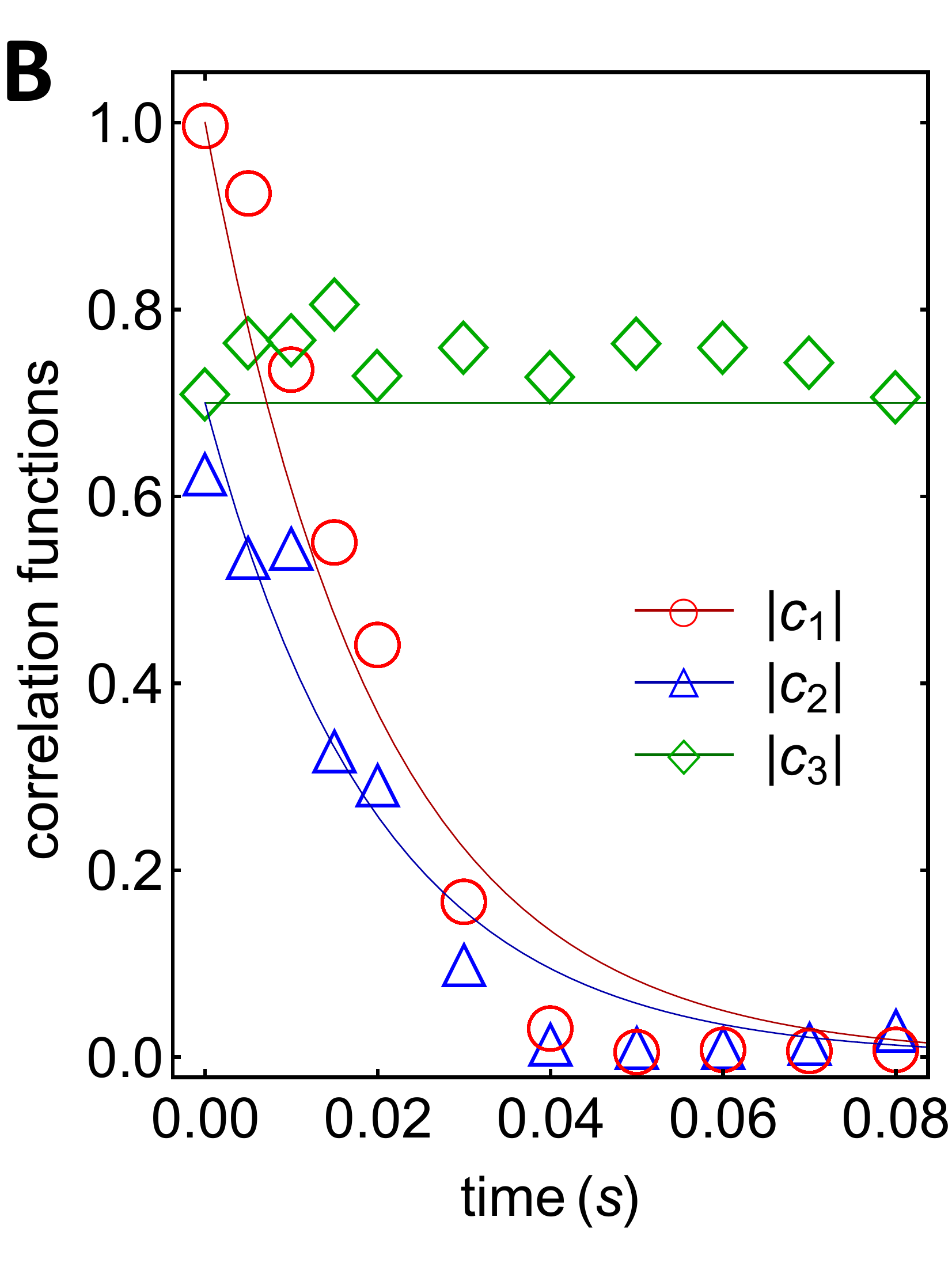} \hspace*{.1cm}
\includegraphics[width=0.164\textwidth]{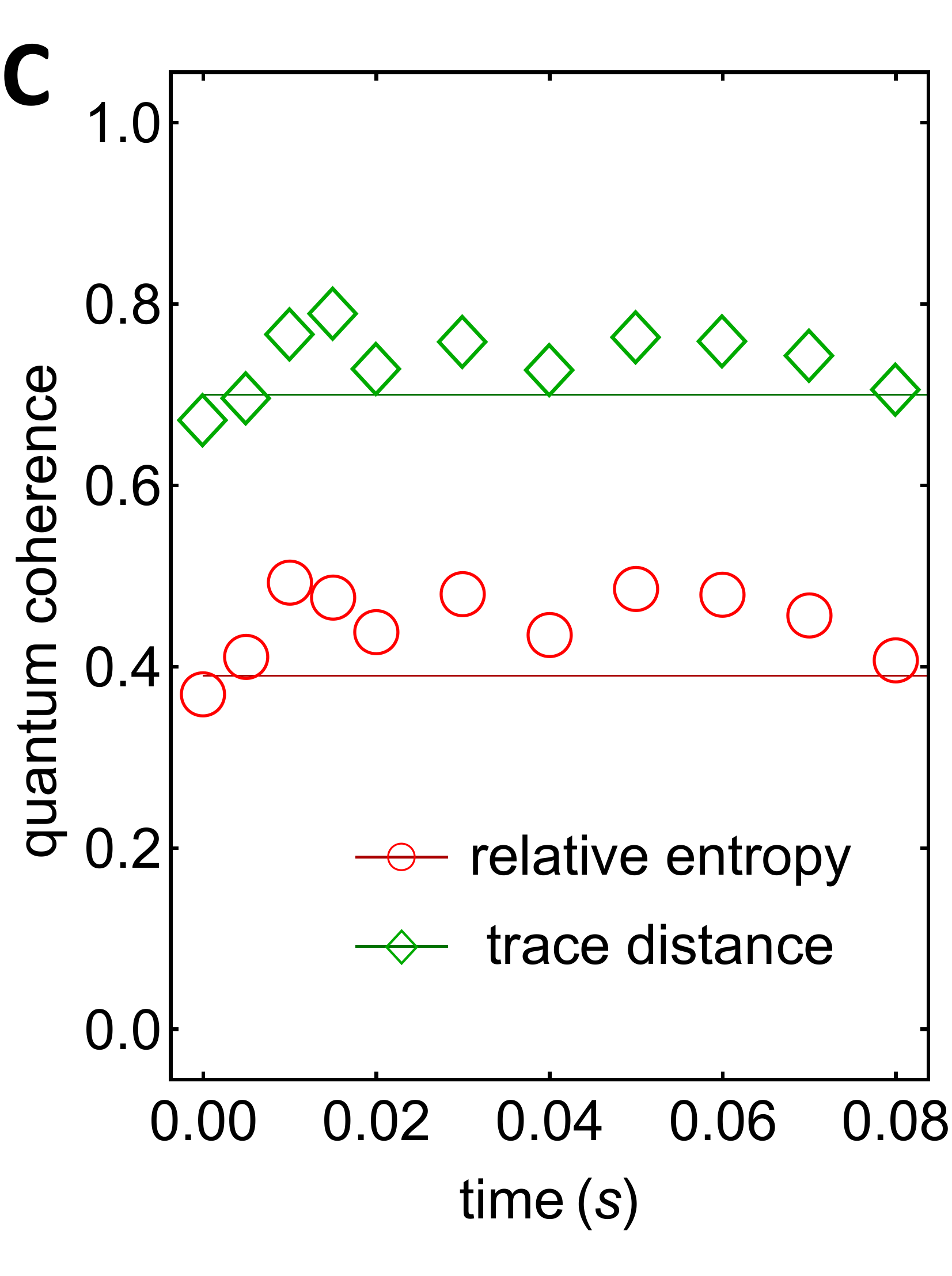}
\caption{\label{fig4q} \footnotesize {\sf \bfseries A}: Pulse sequence (with time flowing from left to right) to prepare four-qubit generalised Bell states encoded in the
$^1$H, $^{19}$F, $^{13}$C, and $^{31}$P nuclear spins of the $^{13}$C$^O$-$^{15}$N-diethyl-(dimethylcarbamoyl)fluoromethyl-phosphonate molecule (whose coupling topology is illustrated as an inset) by an INEPT-like procedure, where $d_{kl}=1/(4J_{kl})$ and $J_{kl}$ is the scalar coupling between spins $k$ and $l$. Light-gray rectangles denote continuous-wave pulses, used to decouple the $^{15}$N nucleus. The dark grey bar denotes a variable pulse, applied to set the desired correlation triple $\{c_j(0)\}$. Thicker (red) and thinner (blue) bars denote $\pi$ and $\pi/2$ pulses, respectively; the phases of the striped $\pi/2$ pulses were cycled to construct each density matrix element. After the preparation stage, the system was  left to decohere in its environment;  $\pi$ pulses were applied in the middle of the evolution to avoid $J_{kl}$ oscillations. The final $\pi/2$ pulses served to produce a detectable NMR signal in the $^1$H spin channel. {\sf \bfseries B}: Evolution of (absolute values of) the correlation functions $|c_j|$ in the experimental states (points), along with  theoretical predictions (solid lines) based on phase damping noise with an effective relaxation time $T_2 \approx 0.04$ s.
{\sf \bfseries C}:
Experimental observation of time-invariant  coherence (in the plus/minus basis) in the four-qubit ensemble, measured by relative entropy (red circles) and normalised trace distance (green diamonds), along with theoretical predictions (solid lines). In panels {\sf \bfseries B}--{\sf \bfseries C}, experimental errors due to pulse imperfections  and coupling instabilities result in error bars within the size of the data points.}
\end{figure*}

Finally, we investigated experimentally the resilience of coherence in  a larger system, composed of four logical qubits. To this aim, we performed a more advanced NMR demonstration
in a BRUKER AVIII $600$ MHz spectrometer equipped with a prototype $6$-channel probe head, allowing full and independent control of up to  $5$ different nuclear spins \cite{Marx2000,marx}. We used the $^{13}$C$^O$-$^{15}$N-diethyl-(dimethylcarbamoyl)fluoromethyl-phosphonate compound \cite{marx}, whose coupling topology is shown in Fig.~\ref{fig4q}{\sf A}. This molecule contains $5$ NMR-active spins ($^1$H, $^{19}$F, $^{13}$C, $^{31}$P and $^{15}$N), therefore we chose to decouple $^{15}$N and encode our four-qubit system in the remaining spins. Each pair of spins $k,l=\{H,F,C,P\}$ were coupled to each other by suitable scalar constants $J_{kl}$ \cite{EPAP}.
We employed an `Insensitive Nuclei Enhanced by Polarization Transfer' (INEPT)-like procedure \cite{inept} to prepare a generalised BD state $\rho(0) = \frac{1}{16}\big(\mathbb{I}^{\otimes 4} + c_1(0)\sigma_1^{\otimes 4} + c_2(0)\sigma_2^{\otimes 4} + c_3(0)\sigma_3^{\otimes 4}\big)$ with initial correlation functions $c_1(0) = 1$, $c_2(0) = c_3(0)= 0.7$ \cite{EPAP}, as detailed in Fig.~\ref{fig4q}{\sf A}. After evolution in a natural phase damping environment as before, the coherence dynamics was measured by a non-tomographic detection method similarly to what was done in \cite{silva}, reading out the correlation triple (see Fig.~\ref{fig4q}{\sf B}) from local spin observables on the ${}^1$H nucleus, whose spectrum exhibited the best signal-to-noise ratio \cite{EPAP}. The results in Fig.~\ref{fig4q}{\sf C} demonstrate, albeit with a less spectacular accuracy than the two-qubit case, time-invariant coherence in the plus/minus basis in our  four-qubit complex, as measured by normalised trace distance and relative entropy of coherence; the latter quantity also coincides with the global discord, a measure of multipartite quantum correlations \cite{Rulli,XuG}, in generalised BD states.

In conclusion, we demonstrated experimentally in two different  room temperature NMR setups that coherence, the quintessential signature of quantum mechanics \cite{AlexRMP}, can resist decoherence under particular dynamical conditions, in principle with no need for external control. While only certain states feature exactly time-invariant coherence in theory, more general states were shown to maintain a guaranteed amount of coherence within the experimental timescales.
These phenomena, here observed for two- and four-qubit ensembles,
are predicted to occur in larger systems composed by an arbitrary (even) number of qubits \cite{Frozen}. It is intriguing  to wonder whether biological systems such as light-harvesting complexes, in which quantum coherence effects persist under exposure to dephasing environments \cite{Engel2007, Panitchayangkoon2010,  Chin2013}, might have evolved towards exploiting natural mechanisms for coherence protection similar to the ideal one reported here; this is a topic for further investigation \cite{Lloyd2011}.

While this Letter realises a proof-of-principle demonstration, our findings can impact on practical applications, specifically on noisy quantum and nanoscale technologies.
In particular, in quantum metrology \cite{Giovannetti2004}, coherence in the plus/minus basis is a resource for precise estimation of frequencies or magnetic fields generated by a Hamiltonian aligned along the spin-$x$ direction. When decoherence with a preferred transversal direction (e.g., phase damping noise) affects the estimation, as in atomic magnetometry \cite{Magneto,KoloPRX}, a quantum enhancement can be achieved by optimising the evolution time \cite{KoloPRL,KoloPRX} or using error correcting techniques \cite{Error1,Error2}. Here we observed instances in which coherence is basically unaffected by transversal dephasing noise. This suggests that the states prepared here (or others in which similar phenomena occur, such as GHZ states; see also \cite{DongF}) could be used as metrological probes with sensitivity immune to decoherence. Furthermore, it has been recently shown that the quantum advantage in discriminating phase shifts generated by local spin-$x$ Hamiltonians is given exactly by the `robustness of coherence' \cite{Napoli2016} in the plus/minus basis, a measure equal to the trace distance of coherence for BD states: this implies that the performance of such an operational task can in principle run unperturbed, if the probes are initialised as in our demonstration, in presence of a natural dephasing environment. We will explore these applications experimentally in future works.

\begin{acknowledgments}%
This work was supported by the European Research Council (Starting Grant No.~637352 GQCOP), the Brazilian funding agencies FAPERJ, CNPq (PDE Grant No.~236749/2012-9), and CAPES (Pesquisador Visitante Especial-Grant No.~108/2012), the Brazilian National Institute of Science and Technology of Quantum Information (INCT/IQ), and the German DFG SPP 1601 (Grant No.~Gl 203/7-2).
\end{acknowledgments}

\bibliographystyle{mybst}
\bibliography{FBibliography}



\clearpage
\addtolength{\textheight}{.3cm}
\appendix

\section*{Supplemental Material}
\section{Experimental details}

\subsection{Two Qubits System}
\subsubsection{NMR setup}
The NMR experiments for two-qubit systems were performed in a Chloroform (CHCl$_3$) sample enriched with $^{13}$C, prepared as a mixture of $100$ mg of $99\%$ $^{13}$C-labelled CHCl$_3$ in $0.7$ mL of $99.8\%$ CDCl$_3$. The $^1$H and $^{13}$C spin-$1/2$ nuclei were associated to the first and second qubits, respectively. This system is described by the Hamiltonian
\begin{eqnarray}
\mathcal{H} = \hbar\omega_HI_z^H + \hbar\omega_CI_z^C + 2\pi\hbar JI_z^HI_z^C,
\end{eqnarray}
where $I^k_z$ is the $z$ component of the spin angular momentum of each nucleus $k=H,C$, and $\omega_k$ is their Larmour frequency. On a Varian $500$ MHz liquid-NMR spectrometer, where the experiments were implemented, it corresponds to $\omega_H/2\pi\approx500$MHz and $\omega_C/2\pi\approx125$MHz, where $J$ represents the weak scalar spin-spin coupling which was  measured at $\approx 215$ Hz.

The experiments were performed at room temperature, so that the high temperature approximation ($k_BT \gg \hbar\omega_L$) guarantees that the thermal equilibrium density operator related to this Hamiltonian, $\rho_0 = e^{-\mathcal{H}/k_BT}/\mathcal{Z}$, can be simplified to
\begin{eqnarray}
\rho_0 \approx \frac{1}{4}\left(\mathbb{I}^{A} \otimes \mathbb{I}^{B} + \epsilon\Delta\rho\right),
\label{rho_nmr}
\end{eqnarray}
where $\mathcal{Z} = \sum_me^{-E_m/k_BT}$ is the associated partition function,  $\Delta\rho = I_z^H + I_z^C/4$ is the so-called deviation matrix, and $\epsilon = \hbar\omega_H/k_BT \sim 10^{-5}$. The application of radiofrequency (rf) pulses and the evolution under spin interactions allow for easy manipulation of the thermal state $\rho_{0}$ in order to produce different states with an excellent control of angle and phase. This procedure is described by the Hamiltonian
\begin{eqnarray}
&&\mathcal{H} =\hbar(\omega_H-\omega_{rf}^H)I_z^H + \hbar(\omega_C-\omega_{rf}^C)I_z^C \\
&&\,\,\,\,-\hbar\omega_1^H(I_x^H\cos\phi^H + I_y^H\sin\phi^H) - \hbar\omega_1^C(I_x^C\cos\phi^C + I_y^C\sin\phi^C) \nonumber
\end{eqnarray}
where $\omega_{rf}^k$ is the frequency of the rf field for nucleus $k$ (on resonance: $\omega_{rf}^k \approx \omega_k$), $\omega_1^k$ is the nutation frequency and $\phi^k$ their respective phase. As only the deviation matrix $\Delta\rho$ is affected by those unitary transformations, it is convenient to write the resulting state as $\rho_{total} = \left[(1-\epsilon)(\mathbb{I}^{A} \otimes \mathbb{I}^{B}) + \epsilon\rho\right]/4$, from which we define the logical NMR density matrix as the state $\rho \equiv \left[(\mathbb{I}^{A} \otimes \mathbb{I}^{B}) + \Delta\rho\right]/4$. The state preparation procedure discussed in the main text refers to engineering $\rho$ into any desired two-qubit BD state.

\subsubsection{Decoherence processes}
NMR naturally provides well characterised environments, characterised by the Phase Damping (PD) and Generalised Amplitude Damping (GAD) channels acting on each qubit \cite{Nielsen2010}. PD is associated to loss of coherence (in the computational basis) with no energy exchange and is specified by the following Kraus operators,
\begin{eqnarray}
K^P_0 = \sqrt{1-\frac{q(t)}{2}}\ \mathbb{I},\ \ \ K^P_3=\sqrt{\frac{q(t)}{2}}\ \sigma_3,
\label{PDkraus}
\end{eqnarray}
where the $q(t)$ damping function is related to the characteristic relaxation time $T_2$ by $q(t) = (1-e^{-t/T_2})$.

On the other hand, the GAD channel is associated to energy exchange between system and environment and can be written in Kraus operator form by \begin{eqnarray}
K^G_0 = \sqrt{p}\left( \begin{array}{cc}
1 & 0 \\
0 & \sqrt{1-u(t)} \end{array} \right),\ \ \  K^G_1 = \sqrt{p}\left( \begin{array}{cc}
0 & \sqrt{u(t)} \nonumber \\
0 & 0 \end{array} \right),\\ K^G_2 = \sqrt{1-p}\left( \begin{array}{cc}
\sqrt{1-u(t)} & 0 \\
0 & 1 \end{array} \right),\ \ \ K^G_3 = \sqrt{1-p}\left( \begin{array}{cc}
0 & 0 \\
\sqrt{u(t)} & 0 \end{array} \right), \nonumber \\
\end{eqnarray}
where $u(t) = 1-e^{-t/T_1}$ and $p=1/2-\alpha$ with $\alpha = \hbar \omega_{L} / 2 k_{B} T$.

As shown in Fig. $1{\sf A}$ of the main text, during the evolution period no refocusing pulses were applied. This implies that the phase damping function $q(t)$ decayed naturally according to the characteristic relaxation time $T_2$. We note that the effective $T_2$ for our experiment depends not only on the thermally induced fluctuations of longitudinal fields (standard so-called  $T_2^{*}$ NMR  contribution) but also on static field inhomogeneities. This dependence makes the PD decoherence process occur faster, guaranteeing that no GAD effects should be effectively observed during the experiment time domain. The characteristic relaxation times were measured as $T_1^H = 7.53$ s, $T_2^H = 0.14$ s, $T_1^C = 12.46$ s, $T_2^C = 0.90$ s, which satisfy $T_1 \gg T_2$, as desired.

\subsubsection{Quantum state tomography}
The quantum state tomography for this two-qubit system was performed applying the simplified procedure proposed in Ref.~\cite{LongQST}. In this case the full matrix reconstruction is obtained after performing the local operations: $II$, $IX$, $IY$, $XX$ on each qubit. Here, $I$, $X$ and $Y$, correspond, respectively, to the identity operation, a $\pi/2$ rotation around the x-axis, and a $\pi/2$ rotation around the y-axis. This set of operations provides a $16\times16$ system of equations, whose solution gives the density matrix elements.

\subsection{Four Qubits System}
\subsubsection{NMR setup}
The four-qubit experiment was performed in a BRUKER AVIII $600$ MHz spectrometer equipped with a prototype $6$-channel probe head as described in the main text, see \cite{marx} for further details. The molecule chosen to encode the $4$-qubit spin system was the $^{13}$C$^O$-$^{15}$N-diethyl-(dimethylcarbamoyl)fluoromethyl-phosphonate compound, whose coupling topology is shown in Fig.~\ref{fig4q}{\sf A}; a detailed description of its synthesis can be found again in \cite{marx}. This molecule contains $5$ NMR-active spins ($^1$H, $^{19}$F, $^{13}$C, $^{31}$P and $^{15}$N). After decoupling $^{15}$N, the Hamiltonian of the remaining four-spin system can be written as
\begin{eqnarray}
\mathcal{H} &=& \hbar(\omega_H-\omega_{rf}^H)I_z^H + \hbar(\omega_F-\omega_{rf}^F)I_z^F  + \hbar(\omega_C-\omega_{rf}^C)I_z^C +  \nonumber \\
&+& \hbar(\omega_P-\omega_{rf}^P)I_z^P + 2\pi\hbar J_{HC}I_z^HI_z^C + 2\pi\hbar J_{HF}I_z^HI_z^F + \\
&+& 2\pi\hbar J_{HP}I_z^HI_z^P+ 2\pi\hbar J_{FC}I_z^FI_z^C + 2\pi\hbar J_{FP}I_z^FI_z^P + 2\pi\hbar J_{PC}I_z^PI_z^C, \nonumber
\end{eqnarray}
where $\omega_k$ is the Larmour frequency of each spin nucleus, measured in $\omega_H/2\pi \approx 600$ MHz, $\omega_F/2\pi \approx 565$ MHz, $\omega_C/2\pi \approx 151$ MHz, $\omega_P/2\pi \approx 243$ MHz.
The term $J_{kl}$ represents the scalar spin-spin coupling constant between spins $k$ and $l$. In this experiment, all density operator components were generated by the $^1$H magnetisation using the interactions of $^1$H with the other nuclei, therefore the relevant $J_{kl}$ couplings were measured, obtaining: $J_{HF} \approx 46.45$ Hz, $J_{HP} \approx 9.64$ Hz, $J_{HC} \approx 3.99$ Hz. As $^1$H presented the spectrum with the best signal-to-noise ratio, all experiments were also recorded in the $^1$H channel.
The experimentally determined transverse relaxation times \cite{marx} were $6.2$ s ($^1$H), $15.4$ s ($^{13}$C), $4.6$ s ($^{19}$F) and $0.2$ s ($^{31}$P). The effective $T_2$ of our observed dynamics was estimated at $0.04$ s.


The initial ${\cal M}^3_4$ (alias generalised BD) state,
\begin{equation}
\rho(0) = \frac{1}{16}\left(\mathbb{I}^{\otimes 4} + c_1(0)\sigma_1^{\otimes 4} + c_2(0)\sigma_2^{\otimes 4} + c_3(0)\sigma_3^{\otimes 4}\right)
,
\end{equation}
was prepared according to the pulse sequence displayed in Fig.~\ref{fig4q}{\sf A} in the main text. Specifically, to distinguish the physical qubits, we associate each $\sigma_i^{\otimes 4}$ term to the spin operators $I_i^HI_i^FI_i^CI_i^P$. Each of these terms is independent of the others and also interacts independently with the environment (considering the same kind of decoherence process described before for the two-qubits system), therefore each term was prepared separately.

Firstly, a continuous-wave (cw) pulse was applied in $^{19}$F,  $^{13}$C, and $^{31}$P to guarantee that all terms would be generated from the $^1$H magnetisation. Then, a variable pulse was applied to $^1$H in order to produce the correct scaling, i.e.~the correlation triple ($c_1(0) = 1$, $c_2(0) = 0.7$, $c3(0) = 0.7$). Each four-qubit term $\sigma_i^{\otimes 4}$ was then achieved by an INEPT-like transfer step. After this block, a $\pi/2$ pulse was applied on the first spin ($^1$H) to produce the $\sigma_z^{\otimes 4}$ term and/or on the other $3$ spins to produce $\sigma_x^{\otimes 4}/\sigma_y^{\otimes 4}$, where the pulse phase was  appropriately changed.
To guarantee the quality of the final state, a phase cycling was introduced in this step. It was designed to preserve the four-qubit coherence term $\sigma_i^{\otimes 4}$ and eliminate the others. This was accomplished with an $8$-step phase cycling corresponding to a $45^{\circ}$ step rotation.

\subsubsection{Evolution and acquisition}

After the state preparation stage, the system was allowed to interact with its natural environment and suffer decoherence in the form of phase damping. During this period, the $J_{kl}$ couplings between some of our spins and $^{15}$N could interfere so a cw decoupling pulse was applied to the latter channel. A refocusing $\pi$ pulse was applied to the other spins in order to account for some inhomogeneity effects and avoid oscillations in the measured signal, due to any other $J_{kl}$  coupling evolutions. Notice that these inter $\pi$ pulse delays were much longer (order of miliseconds) compared to the correlation time of the thermal fluctuations (order of nanoseconds) of the internal fields responsible for the decoherence;  therefore, they did not act as a decoupling field, and the evolution of the sample was still effectively control-free.

The final step amounted to a $\pi/2$ pulse to convert each multiqubit element (like e.g.~$I_x^HI_z^FI_z^CI_z^P$), in a NMR-detectable term on the $^1$H spin channel.
 This method is akin to what was implemented in \cite{silva}, where it was shown that a direct detection is as good as a full state tomography to assess the dynamics of BD states under noise preserving their BD form.

\section{Theoretical details}

\subsection{Geometric quantifiers of coherence and correlations}

A rigorous and general formalism to quantify the coherence of a $d$-dimensional quantum state $\rho$ with respect to a given reference basis $\{|e_i\rangle\}_{i=1}^d$ can be found in \cite{Baumgratz2014} within the setting of quantum resource theories \cite{Brandao2015}. Natural candidates for quantifying coherence arise from a rather intuitive geometric approach, wherein the distance from $\rho$ to the set of states diagonal in the reference basis (known as incoherent states) is considered, provided the adopted distance satisfies the following constraints: contractivity under completely positive trace preserving (CPTP) maps, i.e. $D(\Phi(\rho),\Phi(\tau))\leq D(\rho,\tau)$ for any CPTP map $\Phi$; joint convexity, i.e. $D(\sum_i p_i \rho_i,\sum_i p_i \tau_i)\leq \sum_i p_i D(\rho_i,\tau_i)$ for any probability distribution $\{p_i\}$; plus some additional properties listed in \cite{Vedral1997}. In particular,
\begin{equation}
C_D(\rho)\equiv\inf_{\delta \in \mathcal{I}} D(\rho,\delta),
\end{equation}
with $\mathcal{I}$ being the set of incoherent states, defines full coherence monotones $C_D$ (i.e., satisfying all requirements of the resource theory defined in \cite{Baumgratz2014}) if one chooses for example the following distance functionals: relative entropy distance \cite{Baumgratz2014} $D_{RE}(\rho,\tau) = S(\rho||\tau)$, where $S(\rho||\tau)=\text{Tr}[\rho (\log(\rho)-\log(\tau))]$ is the quantum relative entropy; and fidelity based distance \cite{Streltsov2015} $D_F(\rho,\tau) = 1 - F(\rho,\tau)$, where $F(\rho,\tau)=\left(\text{Tr}\left(\sqrt{\sqrt{\rho} \tau \sqrt{\rho}}\right)\right)^2$ is the Uhlmann fidelity. It is still currently unknown whether the trace distance $D_{\text{Tr}} \left( \rho,\tau \right)=\text{Tr}\left(\sqrt{(\rho-\tau)^2}\right)$ induces a full coherence monotone (in particular, it is still unclear whether property C2b of \cite{Baumgratz2014} is satisfied by such a measure), even though it is both contractive and jointly convex. However, for BD states the trace distance of coherence is equal to the $l_{1}$ norm of coherence (note that we adopt a normalised definition for the trace distance equal to twice the conventional one \cite{Nielsen2010}), which is a full coherence monotone \cite{Baumgratz2014,Frozen}.

Analogously, one can define faithful measures of correlations such as total correlations, discord-type quantum correlations and entanglement in the following way.

For total correlations,
\begin{equation}
T_D(\rho)\equiv\inf_{\pi \in \mathcal{P}} D(\rho,\pi),
\end{equation}
where $\pi = \rho^A\otimes\tau^B$, with $\rho^{A}$ ($\tau^B$) being an arbitrary state of subsystem $A$ ($B$), form the set of product states $\mathcal{P}$.

For quantum correlations \cite{Modi2012},
\begin{equation}\label{eq:QD}
Q_D(\rho)\equiv\inf_{\chi \in \mathcal{C}} D(\rho,\chi),
\end{equation}
where $\chi=\sum_{ij} p_{ij} |i^A\rangle\langle i^A|\otimes|j^B\rangle\langle j^B|$, with $\{p_{ij}\}$ being a joint probability distribution and $\{|i^A\rangle\}$ ($\{|j^B\rangle\}$) an orthonormal basis of subsystem $A$ ($B$), form the set of classical states $\mathcal{C}$.

For entanglement \cite{Vedral1997},
\begin{equation}
E_D(\rho)\equiv\inf_{\sigma \in \mathcal{S}} D(\rho,\sigma),
\end{equation}
where $\sigma=\sum_{i}p_{i}\rho_i^A\otimes\tau_i^B$, with $\{p_i\}$ being a probability distribution and $\rho_i^A$ ($\tau_i^B$) arbitrary states of subsystem $A$ ($B$), form the set of separable states $\mathcal{S}$.

Finally, within this unifying distance-based approach, yet in a quite different way, it is also possible to identify the classical correlations of a state $\rho$ as follows:
\begin{equation}
P_D(\rho) = \inf_{\{\chi_\rho:Q_D(\rho)=D(\rho,\chi_\rho)\}}\ \inf_{\pi\in\mathcal{P}} D(\chi_\rho,\pi),
\end{equation}
where the first infimum is taken with respect to the subset of $\mathcal{C}$ formed by all the classical states $\chi_\rho$ which solve the optimisation in Eq.~(\ref{eq:QD}) (i.e., all the classical states which are the closest to $\rho$ according to $D$), and the second infimum corresponds to the distance between each such $\chi_{\rho}$ and the
set of product states $\mathcal{P}$ \cite{Modi2010,Bromley2014,Paula2014}.

\subsection{Conditions for time-invariant coherence}

\subsubsection{Two qubits}

We now outline the general conditions such that constant coherence can be observed for all time \cite{Frozen}, constant quantum correlations can be observed up to a switch time $t^\star$ \cite{Cianciaruso2015}, and constant classical correlations can be observed after the switch time $t^\star$, when considering two-qubit BD states undergoing local nondissipative decoherence.

Consider two qubits $A$ and $B$ initially in a BD state:
\begin{equation}\label{eq:BDstateS}
\rho(0)=\frac{1}{4}\left(\mathbb{I}^A\otimes\mathbb{I}^B + \sum_{\alpha=1}^3 c_\alpha(0) \sigma_\alpha^A\otimes\sigma_\alpha^B\right)
\end{equation}
and such that the following special initial condition is satisfied,
\begin{equation}\label{eq:InitialConstraint}
c_i(0)= - c_j(0)c_k(0),
\end{equation}
where $\mathbb{I}$ is the $2\times 2$ identity, $\sigma_\alpha$ is the $\alpha$-th Pauli matrix and $\{i,j,k\}$ is a fixed chosen permutation of $\{1,2,3\}$. Let the two (noninteracting) qubits undergo local Markovian flip-type decoherence channels towards the $k$-th spin direction (i.e. bit-flip noise for $k=1$, bit-phase-flip noise for $k=2$, and phase-flip noise for $k=3$).
Their evolved global state at any time $t$ is then represented by
\begin{equation}
\rho(t)=\sum_{\alpha,\beta=0}^3 K_\alpha\otimes K_\beta \rho(0) K_\alpha^\dagger\otimes K_\beta^\dagger,
\label{Flipkraus}
\end{equation}
where
\begin{equation}
K_0 = \sqrt{1-\frac{q(t)}{2}} \mathbb{I},\ \ \  K_i=0,\ \ \ K_j=0,\ \ \ K_k=\sqrt{\frac{q(t)}{2}} \sigma_k,
\end{equation}
$q(t)=1-e^{-\gamma t}$ is the strength of the noise, $\gamma$ is the decoherence rate, and $\{i,j,k\}$ is the permutation of $\{1,2,3\}$ fixed in the initial condition, Eq.~(\ref{eq:InitialConstraint}). One can easily see that the evolved state is still a BD state, whose corresponding correlation function triple is given by
\begin{equation}\label{supevolve}
c_i(t)= c_i(0) e^{-2\gamma t},\ \ \ c_j(t)= c_j(0) e^{-2\gamma t},\ \ \ c_k(t)=c_k(0).
\end{equation}

Focusing first on coherence, in \cite{Frozen} it was shown that, according to any contractive and jointly convex distance, one of the closest incoherent states $\delta_{\rho(t)}^{(\alpha)}$ to the evolved BD state $\rho(t)$, with respect to the basis consisting of tensor products of eigenstates of $\sigma_\alpha$, is just the Euclidean projection of $\rho(t)$ onto the $c_\alpha$-axis, i.e. it is still a BD state and its triple is given by $\{\delta_{\alpha\beta} c_\alpha(t)\}_{\beta=1}^3$, for any $\alpha\neq i$. Moreover, in \cite{Bromley2014} it was shown that any contractive distance between the evolved BD state $\rho(t)$ and its Euclidean projection onto the $c_j$-axis must be constant for any $t$. It immediately follows that any valid distance-based measure of coherence $C^{(j)}_D(\rho(t))$ of the evolved state, with respect to the product basis consisting of tensor products of eigenstates of $\sigma_j$, is invariant for any time $t$.

For quantum correlations, according to any contractive and jointly convex distance, one of the closest classical states $\chi_{\rho(t)}$ to the evolved BD state $\rho(t)$ is just the Euclidean projection of $\rho(t)$ onto the closest $c$-axis, with triple given by $\{\delta_{\alpha\beta} c_\alpha(t)\}_{\beta=1}^3$ and $\alpha$ set by $| c_{\alpha}(t) | = \max \{|c_{\beta}(t)|\}_{\beta=1}^{3}$ \cite{Cianciaruso2015}. When $|c_{j}(0)| > |c_{k}(0)|$ then $\alpha = j$ until the switch time $t^{\star}=\frac{1}{2\gamma} \ln \left|\frac{c_{j}(0)}{c_{k}(0)}\right|$, with $\alpha = k$ afterwards. Combined with the result in \cite{Bromley2014} that any contractive distance between the evolved BD state $\rho(t)$ and its Euclidean projection onto the $c_j$-axis must be constant for any $t$, this immediately implies time-invariance of quantum correlations up until the switch time $t^{\star}$. No general proof has yet been found for the subsequent time-invariance of classical correlations after the switch time $t^\star$ for any contractive and jointly convex distance, but this has been observed in particular cases (based e.g.~on relative entropy, trace and fidelity based distances) \cite{Mazzola2010,Bromley2014,Aaronson2013}. The time-invariance of classical correlations is related to the finite-time emergence of the pointer basis during the dynamics \cite{Cornelio2012,Paula2013}.

We note that in our two-qubit experiment we have implemented exactly an instance of the above conditions, specifically in the case $i=2$, $j=1$ and $k=3$; the corresponding phase-flip noise reduces precisely to the PD channel occurring in NMR, as it can be seen by comparing the Kraus operators in Eqs.~(\ref{PDkraus}) and (\ref{Flipkraus}). The decoherence rate in our demonstration was given by $\gamma = \frac{T_2^H+T_2^C}{2 T_2^HT_2^C}$.
With this evolution, the switch time is obtained as $t^\star = \frac{1}{2\gamma} \ln\left|\frac{c_1(0)}{c_3(0)}\right|$. For an initial BD state with $c_1(0) = 1$, $c_2(0) = 0.7$ , $c_3(0) = -0.7$ as we prepared, respecting the constraint in Eq.~(\ref{eq:InitialConstraint}), the expected switch time was $t^\star \approx 0.043$ s, which was found in excellent agreement with the experimental data. Time-invariant coherence in the plus/minus basis (i.e. the eigenbasis of $\sigma_1$) was observed according to any known valid geometric measure $C_D$ (Fig.~1 of the main text). Entanglement and total correlations instead decay monotonically without experiencing any interval of time-invariance in the considered dynamical conditions.

\subsubsection{Even $N$ qubits}

BD states are particular instances of a more general class of $N$-qubit states, that we may call generalised BD states or  $\mathcal{M}^3_N$ states \cite{Frozen}, having all maximally mixed marginals and characterised only by the three correlations functions $c_\alpha=\langle \sigma_{\alpha}^{\otimes N} \rangle$, with $j=1,2,3$.
For any even $N \geq 2$, consider $N$ qubits initially in an $\mathcal{M}^3_N$ state,
\begin{equation}\label{eq:BDstateSN}
\rho(0)=\frac{1}{2^{N}}\left( \mathbb{I}^{\otimes N} + \sum_{\alpha=1}^{3} c_{\alpha}(0) \sigma_{\alpha}^{\otimes N}\right),
\end{equation}
and such that the following  initial condition is satisfied,
\begin{equation}\label{eq:InitialConstraintN}
c_{i}(0)=(-1)^{N/2}c_{j}(0)c_{k}(0),
\end{equation}
where  $\{i,j,k\}$ is a fixed chosen permutation of $\{1,2,3\}$.
Let the $N$ (noninteracting) qubits undergo local Markovian flip-type decoherence channels towards the $k$-th spin direction as before (notice that such a dynamics is strictly incoherent \cite{Baumgratz2014,Winter2016} with respect to any product basis $\{\ket{m}\}^{\otimes N}$, with $\{\ket{m}\}$ being the eigenbasis of any of the three canonical
Pauli operators $\sigma_{m}$ on each qubit). The evolved global state of the $N$ qubits at any time $t$ is then represented by
\begin{equation}
\rho(t)=\!\!\!\!\sum_{\alpha_{1},\ldots,\alpha_{N}=0}^{3}\big(K_{\alpha_{1}}(t)\otimes\ldots\otimes K_{\alpha_{N}}(t)\big)\rho(0)\big(K_{\alpha_{1}}(t)\otimes\ldots\otimes K_{\alpha_{N}}(t)\big)^{\dagger}.\label{FlipkrausN}
\end{equation}
Once again, the evolved state is still a ${\cal M}^3_N$ state, whose corresponding correlation function triple is given by Eq.~(\ref{supevolve}).
Then, for any even $N \geq 2$, any valid distance-based measure of coherence $C^{(j)}_D(\rho(t))$ of the evolved state, with respect to the product basis consisting of tensor products of eigenstates of $\sigma_j$, is invariant for any time $t$ \cite{Frozen}.

We note that in our four-qubit experiment ($N=4$) we have implemented precisely an instance of the above conditions, specifically in the case $i=2$, $j=1$ and $k=3$, and for an initially prepared $\mathcal{M}^3_4$ state with $c_1(0) = 1$, $c_2(0) = 0.7$, and $c_3(0) = 0.7$, respecting the constraint in Eq.~(\ref{eq:InitialConstraintN}). Time-invariant coherence in the plus/minus basis (i.e. the eigenbasis of $\sigma_1$) was observed according to various geometric measures  $C_D$ (Fig.~3 of the main text).

\subsection{Coherence lower bound from generalised BD states}

We now prove that the coherence of an arbitrary $N$-qubit state $\rho$, with correlation functions $c_j=\langle \sigma_{j}^{\otimes N} \rangle$, is lower bounded by the coherence of the $\mathcal{M}^3_N$ state defined by the same correlation functions. This holds regardless of the number of qubits $N$ and when considering the basis consisting of tensor products of eigenstates of $\sigma_j$, for any $j=1,2,3$ (e.g. the plus/minus basis when $j=1$, as demonstrated experimentally in Fig.~2 of the main text for $N=2$). Such a proof relies on the following two results.

First, as shown in \cite{Cianciaruso2015Entanglement}, any $N$-qubit state $\rho$ can be transformed into an $\mathcal{M}^3_N$ state with the same correlation functions $c_j=\langle \sigma_{j}^{\otimes N} \rangle$ through the map $\Theta$ defined as follows:
\begin{equation}
\Theta(\rho)= \frac{1}{2^{2(N-1)}}{\sum}_{j=1}^{2^{2(N-1)}} U_{j}' \varrho U_{j}'^{\dagger}
\end{equation}
where $U_{j}'$ are the following single-qubit local unitaries
\begin{eqnarray}\label{eq:localunitariesUprime}
\{U_{j}'\}_{j=1}^{2^{2(N-1)}}&=&
\left \{
  \mathbb{I}^{\otimes N},\,
  \{U_{j_{1}}\}_{j_{1}=1}^{2(N-1)},\,
  \{U_{j_{2}}U_{j_{1}}\}_{j_{2}>j_{1}=1}^{2(N-1)},\,
  \cdots \right. \\
&& \left. \cdots  \{U_{j_{2(N-1)}} \ldots U_{j_{2}}U_{j_{1}}\}_{j_{2(N-1)}>\ldots> j_{2}>j_{1}=1}^{2(N-1)}
\right \}, \nonumber
\end{eqnarray}
with
\begin{eqnarray}\label{eq:localunitariesU}
\{U_{j}\}_{j=1}^{2(N-1)}&=&\left\{(\sigma_{1} \otimes \sigma_{1} \otimes I^{\otimes N-2}),
(I \otimes \sigma_{1} \otimes \sigma_{1} \otimes I^{\otimes N-3}),
\ldots, \nonumber\right. \\
&& \left. (I^{\otimes N-3} \otimes \sigma_{1} \otimes \sigma_{1} \otimes I),
(I^{\otimes N-2} \otimes \sigma_{1} \otimes \sigma_{1}), \nonumber \right.\\
&& \left.(\sigma_{2} \otimes \sigma_{2} \otimes I^{\otimes N-2}),
(I \otimes \sigma_{2} \otimes \sigma_{2} \otimes I^{\otimes N-3}),
\ldots, \nonumber \right. \\
&& \left. (I^{\otimes N-3} \otimes \sigma_{2} \otimes \sigma_{2} \otimes I),
(I^{\otimes N-2} \otimes \sigma_{2} \otimes \sigma_{2})
\right\}. \nonumber \\
\end{eqnarray}

Second, as shown below, the map $\Theta$ is an incoherent operation with respect to the basis consisting of tensor products of eigenstates of $\sigma_j$, for any $j=1,2,3$. An incoherent operation is a CPTP map that cannot create coherence, i.e. with Kraus operators $\{K_{i}\}$ satisfying $K_{i} \mathcal{I} K_{i}^{\dagger} \subset \mathcal{I}$ for all $i$, with $\mathcal{I}$ the set of incoherent states with respect to the chosen reference basis. Since any coherence monotone must be non-increasing under incoherent operations \cite{Baumgratz2014}, it follows that the coherence of $\Theta(\rho)$ is less than or equal to the corresponding coherence of $\rho$.

In what follows we will prove that $\Theta$ is an incoherent operation with respect to the basis consisting of tensor products of eigenstates of $\sigma_1$ (i.e. the plus/minus basis), although analogous proofs hold when considering the other two Pauli operators. Since the Kraus operators of the map $\Theta$ are given by $K_j = \frac{1}{2^{N-1}} U_j' $, in order for $\Theta$ to be an incoherent operation in the plus/minus basis we need that $U_j' \delta U_{j}'^{\dagger}\in\mathcal{I}$ for any $\delta\in\mathcal{I}$ and any $j\in\{1,\cdots,2^{2(N-1)}\}$, where $\mathcal{I}$ is the set of states diagonal in this basis. This obviously holds for $j=1$, being $U_1'$ the identity. On the other hand, as it can be seen from Eq.~(\ref{eq:localunitariesUprime}), all the other single-qubit unitaries $U_j'$ are just products of the single-qubit unitaries $U_j$ listed in Eq.~(\ref{eq:localunitariesU}), so that we just need to prove that $U_j\delta U_j^\dagger\in\mathcal{I}$ for any $\delta\in\mathcal{I}$ and for any $j\in\{1,\cdots,2(N-1)\}$. For any $j\in\{1,\cdots,N-1\}$, $U_j$ just leaves any state which is diagonal in the plus/minus basis invariant, it being a tensor product between two $\sigma_1$'s acting on two neighbouring qubits and the identity acting on the remaining ones. Otherwise, for any $j\in\{N,\cdots,2(N-1)\}$, $U_j$ is the tensor product between two $\sigma_2$'s acting on two neighbouring qubits and the identity on the rest of the qubits. Consequently, by using $\sigma_2|\pm\rangle\langle\pm|\sigma_2 = |\mp\rangle\langle\mp|$ and the fact that the general form of a state $\delta$ diagonal in the plus/minus basis is $\delta=\sum_{j_1,j_2,\cdots,j_N=\pm} p_{j_1,j_2,\cdots,j_N} |{j_1,j_2,\cdots,j_N}\rangle\langle{j_1,j_2,\cdots,j_N}|$, we have that, when e.g. $j=N$, then
$U_N \delta U_N^\dagger = \sum_{j_1,j_2,\cdots\!,j_N=\pm} p_{j_1,j_2,\cdots\!,j_N} U_N|{j_1,j_2,\cdots\!,j_N}\rangle\langle{j_1,j_2,\cdots\!,j_N}|U_N^\dagger = \sum_{j_1,j_2,\cdots\!,j_N=\pm} p_{j_1,j_2,\cdots\!,j_N} |{\pi(j_1),\pi(j_2),\cdots\!,j_N}\rangle\langle{\pi(j_1),\pi(j_2),\cdots\!,j_N}|$,
where $\pi(\pm) \equiv \mp$, so that $U_N \delta U_N^\dagger\in\mathcal{I}$. Analogously, one can see that all the remaining single-qubit local unitaries $U_j$ are such that $U_j\delta U_j \in\mathcal{I}$, thus completing the proof.

\clearpage

%
%

\end{document}